\documentclass[a4paper]{article}

\usepackage{INTERSPEECH2020}

\makeatletter
\def\endthebibliography{%
  \def\@noitemerr{\@latex@warning{Empty `thebibliography' environment}}%
  \endlist
}
\makeatother

\title{Early Stage LM Integration Using Local and Global Log-Linear Combination}
\name{Wilfried Michel, Ralf Schl\"uter, Hermann Ney}
\address{
  Human Language Technology and Pattern Recognition, Computer Science Department, \\
  RWTH Aachen University, 52056 Aachen, Germany \\
  AppTek GmbH, 52062 Aachen, Germany}
\email{\{michel,schlueter,ney\}@cs.rwth-aachen.de}

\begin{document}

\maketitle
\begin{abstract}
Sequence-to-sequence models with an implicit alignment mechanism (e.g. attention) are closing the performance gap towards traditional hybrid hidden Markov models (HMM) for the task of automatic speech recognition.
One important factor to improve word error rate in both cases is the use of an external language model (LM) trained on large text-only corpora.
Language model integration is straightforward with the clear separation of acoustic model and language model in classical HMM-based modeling. 
In contrast, multiple integration schemes have been proposed for attention models.

In this work, we present a novel method for language model integration into implicit-alignment based sequence-to-sequence models.
Log-linear model combination of acoustic and language model is performed with a per-token renormalization.
This allows us to compute the full normalization term efficiently both in training and in testing.

This is compared to a global renormalization scheme which is equivalent to applying shallow fusion in training.

The proposed methods show good improvements over standard model combination (shallow fusion) on our state-of-the-art Librispeech system.
Furthermore, the improvements are persistent even if the LM is exchanged for a more powerful one after training.

\end{abstract}
\noindent\textbf{Index Terms}: sequence-to-sequence, language model integration, deep fusion, shallow fusion

\section{Introduction}
For the task of automatic speech recognition (ASR) state-of-the-art approaches rely on statistical models which are estimated on data.
The most valuable kind of data is parallel data with audio signals and the corresponding transcriptions.
We need this kind of data to train our acoustic models (AM).
But unfortunately the amount of parallel data is usually quite limited and it is the most expensive data to obtain.

The amount of available uni-modal data (e.g. audio only or text only) is often several orders of magnitude larger than the amount of parallel data and significant improvements can be obtained by utilizing this additional data.
For audio only data \cite{LongLWZY19} and monolingual data in the translation setting \cite{Backtranslation} the missing labels are created using a pre-trained model and the data can than be added to the available parallel data.
The same can be done for text only data in the ASR setting \cite{rossenbach:ICASSP2020}, but this generally involves a lot of computational effort and constitutes an ill posed problem.

The standard approach for using text only data is to estimate language models, which are then integrated into the speech recognition system.
There are several methods to integrate language models depending on the structure of the acoustic model and the solution to the alignment problem.

\subsection{Language model integration for HMM based models}
The most straightforward approach exists for hidden Markov model (HMM) based models where the need for a language model arises naturally from the Bayes decomposition $P(W|X) = P(X|W)\cdot P(W)/P(X)$ .
During the training of the models the denominator term $P(X)$ usually is ignored, which leads to a decomposition of the estimation problem into two separate instances. 
Language model and acoustic model are then estimated independent of each other.

Further improvements can be obtained by including a pretrained language model into the acoustic model training.
This can be achieved by including $P(X)$ in the training criterion, which leads to the maximum mutual information (MMI) \cite{mmi} training criterion, or by switching to a Bayes risk based training criterion such as minimum phoneme error (MPE) \cite{mpe} or state-level minimum Bayes risk (sMBR) \cite{mbr}.

\subsection{Language model integration for attention models}
For models with an implicit alignment model such as attention, no language model is needed in principle as these models directly output the word posterior probability $P(W|X)$. 
However, it has been shown that including additional text only data is helpful independent of the method used \cite{ToshniwalKCWSL18}.
Commonly used LM integration techniques include \textit{shallow fusion} \cite{GulcehreFXCBLBS15}, \textit{deep fusion} \cite{GulcehreFXCBLBS15}, and \textit{cold fusion} \cite{SriramJSC18}.

Shallow fusion is most similar to the combination of HMM based models as it uses log-linear combination of the word probabilities $P(W|X) = P_{\mathrm{AM}}^\alpha(W|X)\cdot P_{\mathrm{LM}}^\beta(W)/Z$ in place of the Bayes decomposition. 
The normalization term $Z$ is usually ignored which again leads to a decomposition of the training problem.

Deep and cold fusion both integrate an explicit language model into the acoustic model in using the LM hidden states or outputs as additional features for the acoustic model decoder.
These approaches also include the LM in the training process, but no significant improvements over shallow fusion have been obtained so far.
The reader is referred to \cite{ToshniwalKCWSL18} for an in-depth comparison of these combination approaches.

In this work we propose a novel combination technique that combines the simplicity and mathematical elegance of log-linear model combination with the early training integration of cold or deep fusion.

\section{Local and global renormalization}
\label{sec:algo}
In this work we always use the posterior log probability $p(\tilde{w}_1^N|x_1^T)$ of the reference sentence $\tilde{w}^N$ given the acoustic features $x_1^T$ as the starting point for training criterion:
\begin{equation}
F = \log p(\tilde{w}_1^N|x_1^T) \label{eq:general_criterion}
\end{equation}
Different training procedures will arise by how this probability is being modeled.

\subsection{Standard cross entropy}
In most of the current systems the sentence posterior is decomposed over word positions and then directly modeled by the softmax output of a recurrent decoder network.
\begin{align}
F_{\mathrm{CE}} &= \log p(\tilde{w}_1^N|x_1^T) \\
  &= \log \prod_{n=1}^N p_n(\tilde{w}_n | \tilde{w}_1^{n-1}, x_1^T) \\
  &= \sum_{n=1}^N \log q_{\mathrm{AM},n}(\tilde{w}_n | \tilde{w}_1^{n-1}, x_1^T)
\end{align}
This is often referred to as \emph{cross entropy} training criterion. 
In this case no external language model is present in training and no text-only data can be used.

\subsection{Maximum mutual information (MMI)}
A straightforward way to include an external LM is by log-linear model combination. 
This is usually done only during decoding and then called \emph{shallow fusion}.
In this work, we propose to include the LM also during the training of the AM.
This is similar to what has been proposed for RNN-T models in \cite{Weng19mbr-rnnt}.
\begin{align}
F_{\mathrm{MMI}} &= \log p(\tilde{w}_1^N|x_1^T) \\
                 &= \log \frac{q_{\mathrm{AM}}^\alpha(\tilde{w}_1^N|x_1^T) \cdot q_{\mathrm{LM}}^\beta(\tilde{w}_1^N)}{\sum_{w_1^N} q_{\mathrm{AM}}^\alpha(w_1^N|x_1^T) \cdot q_{\mathrm{LM}}^\beta(w_1^N)} \label{eq:mmi}\\
                 &= \alpha \sum_{n=1}^N \log  q_{\mathrm{AM},n}(\tilde{w}_n | \tilde{w}_1^{n-1}, x_1^T) \notag\\
                 &\phantom{=} + \beta  \sum_{n=1}^N \log  q_{\mathrm{LM},n}(\tilde{w}_n | \tilde{w}_1^{n-1})\\ 
                 &\phantom{=} - \log \sum_{w_1^N} \prod_{n=1}^N q_{\mathrm{AM},n}^\alpha(w_n|w_1^{n-1},x_1^T) \cdot q_{\mathrm{LM},n}^\beta(w_n|w_1^{n-1}) \notag
\end{align}
In decoding we are only interested in the word sequence that maximizes the probability so the denominator is omitted as it does not contribute to the decision.
During training we need the full probability.
The denominator is especially important for training as this is the only place where the external language model enters the gradient of the acoustic model.

The denominator contains a sum over all possible word sequences which is infeasible to compute in practice.
In our experiments, this sum will be approximated by an $n$-best list instead.

In Equation \ref{eq:mmi} two scales ($\alpha$: AM-scale, $\beta$: LM-scale) are introduced.
Unlike in decoding, where only the ratio between these scales is important, here, because of the denominator, also the absolute magnitude matters.

This training criterion looks very similar to the \emph{Maximum Mutual Information} (MMI) training criterion used for sequence discriminative training of hybrid-HMM models \cite{mmi}.
Hence we will call it MMI criterion in the following.

\subsection{Local fusion}
In the previous section we first decomposed the posterior (\ref{eq:general_criterion}) into acoustic and language model and then into different word positions.
Now we do it in the opposite order:
\begin{align}
F_{\mathrm{local}} &= \log p(\tilde{w}_1^N|x_1^T) \\
                   &= \log \prod_{n=1}^N p_n(\tilde{w}_n|\tilde{w}_1^{n-1},x_1^T) \\
                   &= \sum_{n=1}^N \log \frac{q_{\mathrm{AM},n}^\alpha(\tilde{w}_n | \tilde{w}_1^{n-1}, x_1^T) \cdot q_{\mathrm{LM},n}^\beta(\tilde{w}_n | \tilde{w}_1^{n-1}) }{\sum_w q_{\mathrm{AM},n}^\alpha(w | \tilde{w}_1^{n-1}, x_1^T) \cdot q_{\mathrm{LM},n}^\beta(w | \tilde{w}_1^{n-1})} \label{eq:local-fusion}
\end{align}
Here we see, that instead of one sum over \emph{all word sequences} which is in $\mathcal{O}(V^N)$ we have $N$ sums over the vocabulary which is in $\mathcal{O}(V\cdot N)$.
This makes it much more tractable and we can calculate all sums exactly.
We denote this criterion \emph{local fusion}.

We have again two scales ($\alpha$, $\beta$) which are both relevant due to the sum in the denominator.

Also note that in Equation \ref{eq:local-fusion} the probabilities in the denominator are always conditioned on the reference history $\tilde{w}_n^{n-1}$ and not on a separate history $w_1^{n-1}$ as in the MMI criterion (Equation \ref{eq:mmi}).
This implies that the denominator cannot be dropped at decoding time since it will depend on the word sequence we are optimizing over.


\section{Experimental setup}
\label{sec:exp_setup}
We will investigate the proposed methods on the full Libri\-Speech corpus (1000h) \cite{Librispeech}.

The same acoustic model architecture as in our previous works \cite{ZeyerBISN19,rossenbach:ICASSP2020} will be used for all experiments.
The encoder consists of a combination of CNN followed by 6 BLSTM layers with time sub-sampling by max-pooling.
We use MLP-style attention with weight feedback and a decoder with a single LSTM layer.
As input to the encoder we use 40 dimensional MFCC features which are perturbed by a variant of SpecAugment \cite{SpecAugment,Zhou20:ted-lium}.
The output consists of 10k grapheme BPE units.

The training of the network involves an intricate pre-training scheme with gradually increasing layer size and an additional ctc-loss on the encoder outputs.
The complete scheme can be found in \cite{ZeyerBISN19}.
Just like in \cite{rossenbach:ICASSP2020} we will store one checkpoint after the pre-training phase but well before convergence and if not noted otherwise use this as initialization for all further experiments.
For some experiments (notably all MMI related experiments) we continued training from this checkpoint until convergence with the CE criterion and use this converged CE model as initialization for further training.

For our experiments we use one out of two neural language models.
The \emph{LSTM} LM consists of 4 layers with 2k units each.
The \emph{Transformer} LM consists of 24 layers with dimension 1k and uses multiheaded attention (8 heads) as described in \cite{IrieZSN19}.
Both LMs operate on the same 10k BPE vocabulary as the AM.
If not noted otherwise, the LSTM LM has been used.

In all experiments where the acoustic model is augmented by an LM, we have the choice to do local or global renormalization.
For the training this is denoted as MMI in the global case or local fusion in the local case.
For decoding we again have the same choice.
We tried exchanging the normalization techniques but found that using the matching technique gave slight improvements in some cases.
Therefore, we use shallow fusion to decode MMI and CE models and local fusion to decode local fusion models in all experiments.

All Experiments have been performed using our speech recognition toolkits RASR \cite{rybach2011:rasr} and RETURNN \cite{doetsch2017returnn}.

\subsection{Practical guidelines for the MMI criterion}
While the local fusion criterion worked out of the box,
there were some subtleties for the MMI criterion we would like to point out.

Replacing the sum over all word sequences by an $n$-best list is an approximation whose accuracy strongly depends on an appropriate choice of $n$. 
A larger $n$, however, requires increased memory and computation time. 
Here we chose $n=8$, which was the maximum we could fit in our current GPU memory given our current decoder.
Some works suggest sequence training to be possible with even lower beam sizes \cite{Prabhavalkar:mwer_train}.

Similar to what has been found for hybrid HMM sequence training \cite{Vesel2013SeqTr}, we found it imperative to always include the reference transcript in the $n$-best list. 
If the reference transcript is absent from the top $n$ results, the hypothesis with lowest probability is replaced by the reference. 
Otherwise the beam is kept intact. 

The decoding of sequence to sequence models suffers a length bias problem \cite{lengthBias2016}.
While there are some ways around it without modification of the hypothesis scores \cite{Zhou20:robust-beamsearch}, the prevalent solution is applying a length normalization \cite{MurrayC18WMT} or silence penalty to the decoding scores \cite{Hannun0XC19}.
Naturally we must not distort the scores for the MMI criterion.
All experiments are performed with the vanilla AM and LM scores.

\section{Results}
\label{sec:results}
\subsection{Runtime comparison}
It is expected that the more complex training criteria which also involve forwarding through a language model and the computation of the normalization term will take more time per training epoch than the cross entropy criterion.
In Table \ref{tab:speed} we compare the average training time per sub-epoch (50h) using a single GPU.

For the training using the local fusion criterion we always compute the full normalization.
For the MMI training we approximate the sum by an $n$-best list with $n=8$.
Also the batch size had to be reduced greatly for the MMI training which results in longer training times.

\begin{table}[htb]
  \vspace{-2mm}
  \caption{Training time in minutes per sub epoch of 50h for different training criteria and language models used in training.}
  \label{tab:speed}
  \vspace{-2mm}
  \centering
  \begin{tabular}{ c | c | c | c  }
    \toprule
    Criterion & LM & time [min] & slowdown factor  \\
    \midrule
    CE     &  -     & \phantom{1}48  & 1.00  \\
    \midrule
    local  & LSTM   & \phantom{1}51  & 1.06  \\
    fusion & TRAFO  & \phantom{1}49  & 1.04  \\
    \midrule
    MMI    & LSTM   &           249  & 5.26  \\
    \bottomrule
  \end{tabular}
  \vspace{-3mm}
\end{table}
\noindent
We see that the local LM combination does not slow down the training significantly.
The MMI training increases training time by a factor of 5.
We therefore decided to start all MMI related experiments from a fully converged CE model as initialization as it is usually the case for the MMI training of hybrid HMM models \cite{Vesel2013SeqTr}.

\subsection{Local fusion training criterion}
\subsubsection{AM and LM scales}
From Equation \ref{eq:local-fusion} we observe that for local fusion both relative ($\gamma_{rel}$) and absolute scale ($\gamma_{abs}$) of the acoustic and language model influence the result. 
We define these scales as: $\gamma_{abs} = \alpha$ and $\gamma_{rel} = \beta / \gamma_{abs}$.
As a starting point of our tuning experiments we use $\gamma_{rel}=0.35$ which we estimated from decoding experiments. 

\begin{table}[htb]
  \vspace{-2mm}
  \caption{Comparing performance of models trained with local fusion training criterion with LSTM LM and different scales.}
  \label{tab:local-fusion_scale-tuning}
  \vspace{-2mm}
  \centering
  \begin{tabular}{ c | c | c | c  }
    \toprule
     absolute scale & relative scale & \multicolumn{2}{c}{ dev WER [\%]}   \\
     $\gamma_{abs}$ & $\gamma_{rel}$ &  clean        &   other        \\
    \midrule
       1.0          & 0.0\phantom{5} &  2.8          &   7.9          \\
    \midrule
       5.0          & 0.35           &  3.6          &\hspace{-1.5mm}10.4\\
       3.0          & 0.35           &  \textbf{2.6} &   7.6          \\
       2.0          & 0.35           &  \textbf{2.6} &   \textbf{7.3} \\
       1.5          & 0.35           &  2.7          &   \textbf{7.3} \\
       1.2          & 0.35           &  2.7          &   7.4          \\
       1.0          & 0.35           &  2.9          &   8.1          \\
       0.5          & 0.35           &  3.4          &   9.9          \\
    \midrule
       3.0          & 0.5\phantom{5} &  2.8          &   7.7          \\
       3.0          & 0.4\phantom{5} &  2.7          &   7.7          \\
       3.0          & 0.35           &  \textbf{2.6} &   7.6          \\
       3.0          & 0.3\phantom{5} &  2.7          &   7.5          \\
       3.0          & 0.25           &  2.7          &   8.6          \\
    \midrule
       2.0          & 0.5\phantom{5} &  \textbf{2.6} &   7.4          \\
       2.0          & 0.35           &  \textbf{2.6} &   \textbf{7.3} \\
       2.0          & 0.25           &  2.8          &   7.6          \\                     
    \bottomrule
  \end{tabular}
  \vspace{-3mm}
\end{table}

In Table \ref{tab:local-fusion_scale-tuning} we report the result of the scale tuning. 
First we notice that the decoding optimum is also a good choice for training.
For the absolute scale we find an optimum at around $2-3$.
This is in accordance to the ``old rule of thumb'' we know from sequence training of hybrid HMM models, that the training LM scale should be set to 1 and AM scale should be the inverse of the optimum decoding LM scale. \cite{Schluter99}

\subsubsection{Interchangeability of LM}
When we train the AM together with an LM then it learns to work together with the LM at decoding.
Now we wonder if the AM learns to make use of LMs in general or if it is adapted to the specific LM we used in training.
We therefore train our AM together with the LSTM LM and use it together with the transformer LM in decoding.
The results are shown in Table \ref{tab:change_lm}.
\begin{table}[htb]
  \vspace{-2mm}
  \caption{Comparing the effect of using a different LM for training and testing. Models trained with local fusion criterion except for the first block, which is CE.}
  \label{tab:change_lm}
  \vspace{-2mm}
  \centering
  \begin{tabular}{ c | c | c | c  }
    \toprule
                 &            & \multicolumn{2}{c}{ dev WER [\%]}        \\
     training LM & testing LM &  clean        &   other                  \\
    \midrule
                 & -          &  3.9          &            10.6          \\
       -         & LSTM       &  2.8          &  \phantom{1}7.9          \\
                 & TRAFO      &  2.6          &  \phantom{1}7.3          \\              
    \midrule
                 & -          &  4.9          &            12.7          \\
       LSTM      & LSTM       &  2.5          &  \phantom{1}6.9          \\
                 & TRAFO      &  2.3          &  \phantom{1}6.6          \\
    \midrule
       TRAFO     & TRAFO      &  2.4          &  \phantom{1}6.5          \\
    \bottomrule
  \end{tabular}
  \vspace{-3mm}
\end{table}

The first observation of Table \ref{tab:change_lm} is that the transformer LM improves WER compared to the LSTM LM in any case and using the local fusion criterion always improves significantly over the CE baseline.
Using the transformer LM with the local fusion criterion nicely stacks both improvements.

In this simple setting of matched LMs it is indeed possible to exchange the decoding LM without the need for retraining the AM.
We also notice that decoding a local fusion model without LM degrades performance compared to the CE baseline.

\subsubsection{Joint training of AM and LM}
Until now we have only considered the training of the acoustic model while keeping the language model as is.
Within the framework set by the local fusion training criterion it is also possible to jointly train acoustic and language model.
For this we initialized the AM with the converged CE model and started a joint training with the LSTM LM and local fusion criterion.

The result (WER: $3.2\%$ dev-clean, $8.3\%$ dev-other) was much worse than even using the initial CE model together with the initial language model  so that we quickly abandoned this approach.
We attribute the degradation to the fact that the rich amount of text only data used to train the initial model is absent during this training stage so that \emph{catastrophic forgetting} \cite{Robins93a} happens. 

\subsection{MMI training criterion}
\subsubsection{Scale tuning}
Also when using the MMI training criterion the absolute magnitude of AM and LM scale matters.
We would expect similar results as with the local fusion criterion for attention and the MMI criterion for HMM based models.

\begin{table}[htb]
  \vspace{-2mm}
  \caption{Comparing performance of models trained with MMI training criterion with LSTM LM. }
  \label{tab:mmi_scale-tuning}
  \vspace{-2mm}
  \centering
  \begin{tabular}{ c | c | c | c  }
    \toprule
     absolute scale & relative scale & \multicolumn{2}{c}{ dev WER [\%]}   \\
     $\gamma_{abs}$ & $\gamma_{rel}$ &  clean        &   other        \\
    \midrule
       1.0          & 0.0\phantom{5} &  2.8          &   7.9          \\
    \midrule
       5.0          & 0.35           &  2.6          &   7.3          \\
       1.0          & 0.35           &  2.5          &   7.1          \\
       0.5          & 0.35           &  2.5          &   6.9          \\
       0.1          & 0.35           &  \textbf{2.4} &   \textbf{6.6} \\
       0.01         & 0.35           &  2.4          &   6.6          \\
    \bottomrule
  \end{tabular}
  \vspace{-3mm}
\end{table}

In Table \ref{tab:mmi_scale-tuning} we see that at high absolute scales the performance of MMI trained models is comparable to the performance of local fusion trained models.
With reduced absolute scale, however, the performance of MMI models increases while local fusion models suffer from reduced performance.


\subsubsection{Cross entropy smoothing}
Now we investigate if cross entropy smoothing, an invaluable heuristic for hybrid HMM based sequence training \cite{Vesel2013SeqTr}, will be useful in this setting.
For cross entropy smoothing the MMI objective function is linearly interpolated with the cross entropy objective function.
Usually only a small weight (e.g. 10\%) is assigned to the CE criterion.

Here we equivalently introduce an additional scale $\gamma_{den}$ to the denominator part of Equation \ref{eq:mmi}.
A scale of $\gamma_{den}=1$ then corresponds to no smoothing while $\gamma_{den}=0$ would indicate pure cross entropy.

\begin{table}[tbh]
  \vspace{-2mm}
  \caption{Comparing performance of models trained with MMI training criterion with LSTM LM and different cross entropy smoothing settings. Scales were set to $\gamma_{abs}=0.3$, $\gamma_{rel}=0.35$}
  \label{tab:ce-smooth_scale-tuning}
  \vspace{-2mm}
  \centering
  \begin{tabular}{ c | c | c   }
    \toprule
     denominator scale &  \multicolumn{2}{c}{ dev WER [\%]}   \\
     $\gamma_{den}$    &  clean        &   other        \\
    \midrule
       0.0             &  2.8          &   7.9          \\
    \midrule
       0.1             &  2.7          &   7.7          \\
       0.3             &  2.7          &   7.5          \\
       0.6             &  2.6          &   7.3          \\
       0.9             &  2.5          &   7.0          \\
       1.0             &  \textbf{2.4} &   \textbf{6.8} \\
    \bottomrule
  \end{tabular}
  \vspace{-5mm}
\end{table}

As we can see in Table \ref{tab:ce-smooth_scale-tuning} any value lower than $1$ reduces the model performance.
We tried values larger than $1$, but found that the models diverge quickly.
We therefore conclude that CE smoothing is not useful for attention based models.


\subsection{Comparison of training criteria}
In Table \ref{tab:final} we compare the best models we could get with each training criterion.
In this case we also give the word error rate on the test sets.
The MMI criterion clearly outperforms the local fusion criterion, which in turn improves over the CE baseline.

\begin{table}[tbh]
  \caption{Comparing performance of different training criteria. All hyperparameters were optimized individually on the respective dev sets. Numbers are WER[\%]}
  \label{tab:final}
  \vspace{-2mm}
  \centering
  \begin{tabular}{ c | c | c | c | c | c | c }
    \toprule
    init.    &  train.  &   testing &\multicolumn{2}{c|}{dev WER[\%]}& \multicolumn{2}{c}{ test WER[\%]}  \\
    model    &  crit.   &    LM     & clean     & other      & clean  & other      \\
    \midrule
             &  CE      &   LSTM    &  2.8      & 7.9        & 3.0    & 8.4        \\
     pre-    &          &   TRAFO   &  2.6      & 7.3        & 2.8    & 7.8        \\
             \cmidrule{2-7}
     train   &  local   &   LSTM    &  2.6      & 7.3        & 2.9    & 7.8        \\
             &          &   TRAFO   &  2.3      & 6.4        & 2.6    & 6.9        \\
    \midrule
             &  local   &   LSTM    &  2.5      & 6.9        & 2.8    & 7.6        \\
      con-   &          &   TRAFO   &  2.3      & 6.5        & 2.6    & 6.9        \\
             \cmidrule{2-7}
      verged &  MMI     &   LSTM    &  2.4      & 6.6        & 2.6    & 7.0        \\
             &          &   TRAFO   & \textbf{2.2} & \textbf{6.1} & \textbf{2.3}& \textbf{6.4}\\
    \bottomrule
  \end{tabular}
  \vspace{-6mm}
\end{table}

\section{Conclusion}
\label{sec:conclusion}
In this work we proposed two methods to include additional text only data into the training of attention based implicit alignment models in form of an external trained language model.
Both methods improve over the cross entropy baseline with shallow fusion at decoding time.

The \emph{MMI criterion} leads to the largest improvement of $18\%$ relative on test-other, but at the cost of a five-fold increase in training time. 
This criterion can be applied in a late training stage for fine-tuning of a pretrained model.

The \emph{local fusion criterion} improves over the baseline by $10\%$ rel. on test-other without requiring additional effort or resources.
We see no reason to not adapt to this training criterion immediately.
\vspace{-1mm}

\section{Acknowledgments} 
This work has received funding from the European Research Council (ERC) under the European Union's Horizon 2020 research and innovation programme (grant agreement No 694537, project "SEQCLAS") and from a Google Focused Award. 
The work reflects only the authors' views and none of the funding parties is responsible for any use that may be made of the information it contains. 
The GPU cluster used for the experiments was partially funded by Deutsche Forschungsgemeinschaft (DFG) Grant INST 222/1168-1. 
Simulations were partially performed with computing resources granted by RWTH Aachen University under project nova0003.

The authors would like to thank Albert Zeyer and Wei Zhou for many helpful discussions and Kazuki Irie for providing the pretrained language models.


\bibliographystyle{IEEEtran}

\bibliography{refs}

\end{document}